\documentclass[%
reprint,
amsmath,amssymb,
prstab,
]{revtex4-2}

\usepackage{graphicx}
\usepackage{dcolumn}
\usepackage{bm}
\usepackage{hyperref}
\usepackage[mathlines]{lineno}

\begin{document}
\title{Optimized laser-assisted electron injection into a quasi-linear plasma wakefield}

\author{V. Khudiakov$^1$}
\author{A. Pukhov$^1$}
\address{$^1$Institut für Theoretische Physik I, Heinrich-Heine-Universität Düsseldorf, 40225 Düsseldorf, Germany}
\date{\today}

\begin{abstract}
	We present a novel electron injection scheme for plasma wakefield acceleration. The method is based on recently proposed technique of fast electron generation via laser-solid interaction: a femtosecond laser pulse with the energy of tens of mJ hitting a dense plasma target at $45^o$ angle expels a well collimated bunch of electrons and accelerates these close to the specular direction up to several MeVs. We study trapping of these fast electrons by a quasi-linear wakefield excited by an external beam driver in a surrounding low density plasma. This configuration can be relevant to the AWAKE experiment at CERN. We vary different injection parameters: the phase and angle of injection, the laser pulse energy. An approximate trapping condition is derived for a linear axisymmetric wake. It is used to optimise the trapped charge and is verified by three-dimensional particle-in-cell simulations. It is shown that a quasi-linear plasma wave with the accelerating field $\sim$~2.5~GV/m can trap electron bunches with $\sim$~100~pC charge, $\sim$~60~$\mu$m transverse normalized emittance and accelerate them to energies of several GeV with the spread $\lesssim$ 1\% after 10 m.
\end{abstract}

\maketitle

\section{Introduction}
Plasma wakefield acceleration (PWFA) is an actively developing research area \cite{RMP81-1229,RAST9-63,RAST9-85,RMP90-035002}. It is considered as a promising alternative to traditional methods due to the ability of plasma to provide acceleration gradients $\sim$10-$10^2$~GV/m, which are several orders of magnitude higher than those in metallic structures. In a general PWFA scheme, the plasma wave (called the plasma wakefield) is excited by a relativistic charged particle bunch (called a driver) and propagates in the medium with the phase velocity equal to the velocity of the driver. A witness bunch can be injected into an appropriate phase of this wave and accelerated up to high energies.

Despite the successful proof-of-concept experiments, the main problem of PWFA schemes remains to reach the witness bunch quality close to that of beams produced in the conventional facilities. One of the key factors determining the ultimate parameters of the accelerated bunch is the injection scheme. Significant efforts are made in this direction in order to design a scheme, which provides a well-controlled and stable injection of the witness bunch. The on-axis injection of a pre-accelerated high quality electron bunch \cite{inj-ext1, inj-ext2} usually is the best option. Yet, for some PWFA schemes the on-axis external injection is difficult to implement technically, or the self-trapping from the background plasma may lead to an even higher quality of the witness bunche. The self-trapping in PWFA can be based on the plasma density transition \cite{inj-dens1, inj-dens2, inj-dens3} or an additional field-ionisation within the plasma wave \cite{inj-ion1, inj-ion2}. The trojan-horse injection based on ionization injection in the blow-out regime \cite{trojan} has the potential to generate witness bunches with nm-scale emittances.

Here, we introduce a new method of electron injection via laser-solid interaction. In this scheme, a dense target is placed in low-density plasma on a way of the driver. A laser pulse synchronized with the driver hits the surface of the target at the 45 degree angle. The laser expels electrons from the solid target and accelerates them up to MeV energies in a direction close to that of the reflected laser pulse (the specular direction) \cite{mev-e}. Electrons that catch the appropriate phase of the wakefield can be trapped and further accelerated. We perform 3d particle-in-cell (PIC) simulations of this process. The simulations are split in two parts. In the first part we simulate the laser interaction with the solid target. Thus, we repeat the results of \cite{mev-e} using full 3d PIC simulations. After the laser-solid interaction is over, we store the full phase space of the generated hot electrons. In the second part of the simulations, we inject these hot electrons into a quasi-linear plasma wakefield generated in the low density plasma that surrounds the solid target. The low density plasma parameters and the wakefield amplitude are chosen close to the envisioned parameters of the AWAKE RUN2 experiment \cite{awake}. However, we do not simulate the full self-modulated proton bunch that should drive the plasma wave in the real AWAKE setup. Rather, we employ a "toy model" \cite{toy-model}, where the wakefield is excited by an artificially short proton bunch. The 3d PIC simulations are done using the VLPL code \cite{vlpl}.

The paper is organized as follows. In Sec.~\ref{sec:laser-solid}, we overview the recently proposed scheme of MeV electrons generation via laser-solid interaction and present results of 3d PIC-simulations. In Sec.~\ref{sec:conditions}, we discuss the structure of axially symmetric quasi-linear wakefield and optimal injection parameters for the electrons generated on the first stage. An approximate trapping condition is derived and applied for optimization. In Sec.~\ref{sec:simulation}, we present the results of 3d PIC-simulations of electron bunch acceleration in a quasi-linear wakefield for different injection parameters. The Sec.~\ref{sec:conclusion} concludes the main results.
\section{MeV electrons generation}\label{sec:laser-solid}
\begin{figure}[t!]
	\centering
	\includegraphics[width=\linewidth]{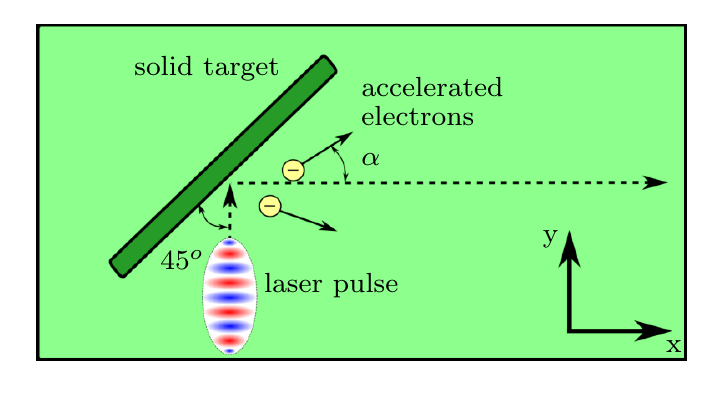}
	\caption{Scheme of laser-solid interaction.}\label{fig0-scheme}
\end{figure}

In the recent paper \cite{mev-e}, an efficient method for electron acceleration up to MeV energies via laser-solid interaction has been demonstrated experimentally and with 2d3v PIC simulations.
According to \cite{mev-e}, the fast electrons are generated via breaking of plasma waves excited by parametric instabilities near the laser reflection point. These electrons are further accelerated by the mechanism of direct laser acceleration\cite{dla}. The scheme is as follows (Fig.~\ref{fig0-scheme}): a laser prepulse ionizes the solid target and forms a plasma layer with a gradient of several laser wavelengths. Next, the main laser pulse hits the target surface at 45 degree angle and accelerates electrons in a direction close to but not exactly coinciding with the pulse reflection.

\begin{figure}[t!]
	\centering
	\includegraphics[width=\linewidth]{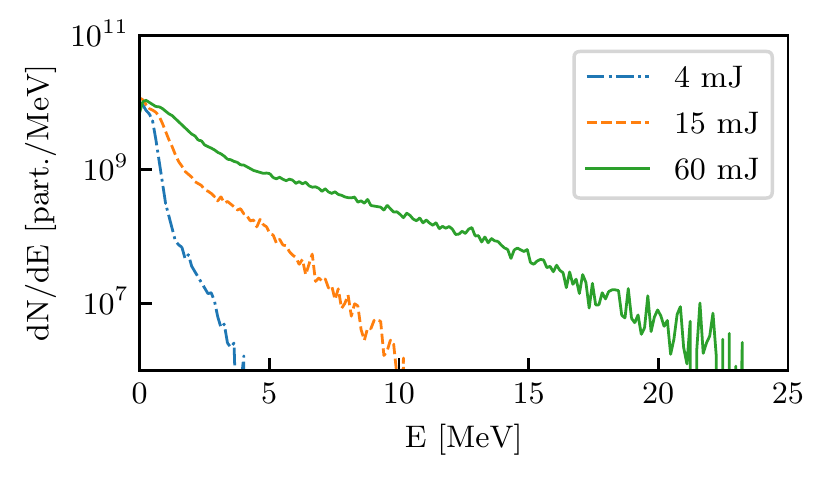}
	\caption{Initial energy spectra depending on laser pulse energy.}\label{fig1-init_spec}
\end{figure}
Here we reproduce the experimental results of \cite{mev-e} using full 3d particle-in-cell simulations with the VLPL code. 
This is the first stage of two-step simulation. In the second stage, the obtained accelerated electrons are injected into the wakefield driven in the surrounding low density plasma and accelerated over the distance of $\sim$10 m. The two separate simulations are unavoidable because of the huge disparity of the time and spacial scales of the physical processes at these two stages. The spatial scale differs by some three orders of magnitude: the laser wavelength equals 0.8 $\mu$m, while the plasma wavelength is $0.126$ cm. It would be numerically inefficient and expensive to resolve the laser wavelength during the whole process.

For the surface of solid target, we utilize the density profile used in \cite{mev-e}:
\begin{equation}
	n_{\xi}/n_c = 4\cdot10^{-7} + 20\cdot e^{-(\xi/(2.5\lambda_0))^4}.
\end{equation}
Here $\xi = (x - y)/\sqrt{2}$ is a coordinate along the surface, $\lambda_0 = 0.8\,\mu m$ is the laser pulse wavelength, $n_c = \pi m_e c^2/(\lambda_0^2e^2)$ is the critical density, $m_e$ is the electron mass, $e$ is the elementary charge, $c$ is the speed of light. The laser pulse has the Gaussian shape with the  following parameters: the duration is fixed to 50 fs, the pulse energy varies in the range of 4-60 mJ, the focal spot diameter $\sigma_{FWHM} = 4\,\mu m$. The simulation domain for this part is $L_x\times L_y\times L_z = 74\lambda_0\times 35\lambda_0\times30\lambda_0$. It is sampled by a grid with the step sizes $h_x\times h_y\times h_z = 0.005\lambda_0\times0.01\lambda_0\times0.01\lambda_0$. We performed three different simulations for the laser pulse energies 4 mJ, 15 mJ, 60 mJ. These pulse energies correspond to the normalized vector potentials $a_0 =  0.5,~1,~2$.

\begin{figure}[t!]
	\centering
	\includegraphics[width=\linewidth]{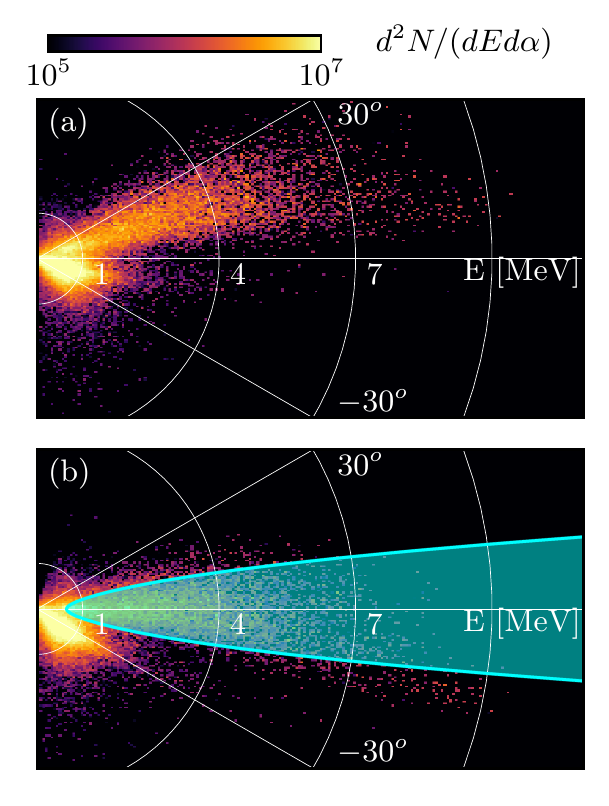}
	\caption{(a) Initial energy-angle distribution (for laser pulse energy 15 mJ). (b) Distribution rotated at optimal angle. Blue region corresponds to electrons under trapping condition.}\label{fig2-en_ang}
\end{figure}

The accelerated electrons have exponential spectra with the effective "temperatures" varying from several MeVs to tens of MeV depending on the laser pulse amplitude (Fig.~\ref{fig1-init_spec}). The energy-angle polar diagrams for these electrons are shown in Fig.~\ref{fig2-en_ang}(a), where the radius corresponds to the electron energy and the angle is measured from the laser reflection direction ($x$-axis). One observes well collimated beams, which are however not exactly aligned with $x$-axis (the specular direction).

\section{Optimal injection conditions}\label{sec:conditions}
We are aiming at maximization of trapped charge and in this scheme we have not so many parameters to vary. These are the phase of the wave in which we inject the hot electrons generated by the laser pulse and the angle between the driver propagation direction and the surface of the solid target.

\begin{figure}[t!]
	\centering
	\includegraphics[width=\linewidth]{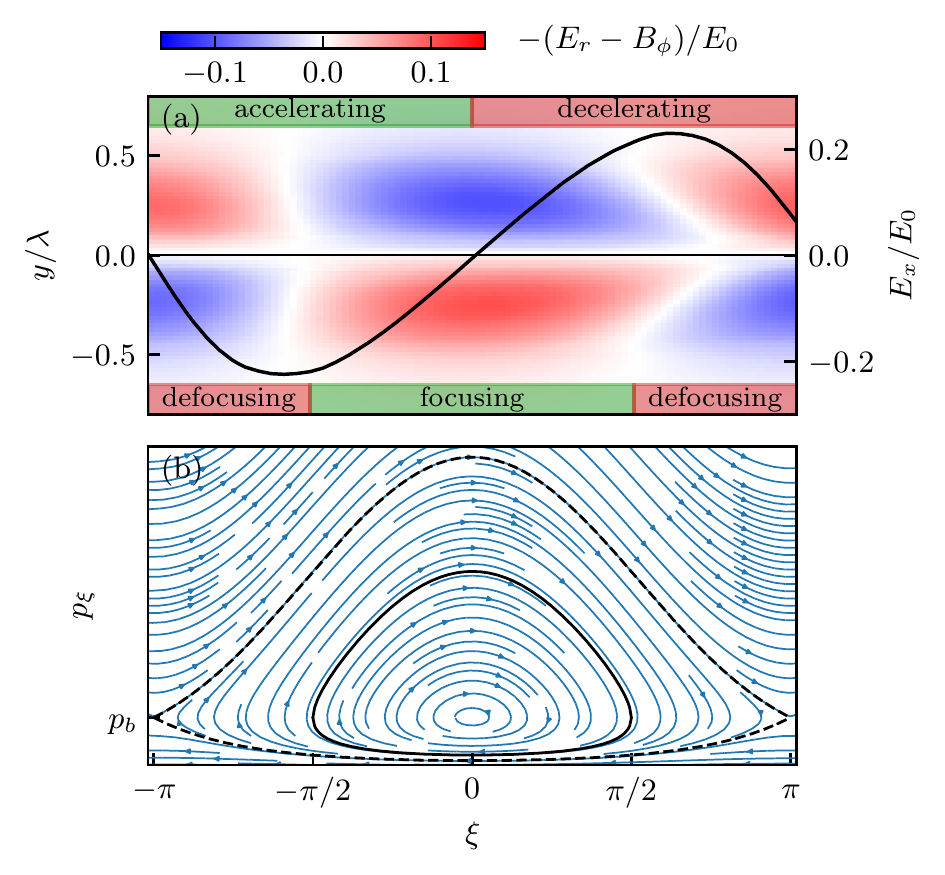}
	\caption{(a) Map of radial force acting on electrons. Black line shows the axial longitudinal electric field $E_x$. (b) Phase portrait for trajectories of test electrons moving in one-dimensional sinusoidal field. The dashed line marks the separatrix for the 1d trapping condition, solid line marks the 2d trapping condition.}\label{fig3-phase}
\end{figure}

From the symmetry considerations, the injection point must be located at the driver propagation axis, as the accelerating fields have a maximum here. Second, the injection must take place when the longitudinal electric field is zero ($\xi = 0$, Fig.~\ref{fig3-phase}(a)), in the beginning of the acceleration phase. As the electrons are initially much slower than the plasma wave, this configuration provides the maximal acceleration path before the electrons arrive into the defocusing region ($\xi < -\pi/2$, Fig.~\ref{fig3-phase}(a)). This is verified numerically in Section 4.

The other important parameter is the angle between the reflected laser and the driver. The laser-generated electrons are reasonably well collimated, but their mean momentum is not exactly aligned with the laser reflection direction (Fig.~\ref{fig2-en_ang}(a)). Therefore, the electrons have an unnecessary mean transverse momentum, which leads to the asymmetry in the $yz$-plane, growth of emittance along the $y$-axis and excessive loss of the electrons in the transverse direction. In order to improve the trapping, the laser-solid setup can be rotated as a whole in the $xy$-plane  to align the average direction of the hot electrons with the low density wakefield axis. As the energy-angle distribution (Fig.~\ref{fig2-en_ang}(a)) is not symmetric, there is no obvious condition for the optimal rotation angle. We need to find a trapping region on the energy-angle diagram and by rotating electron distribution determine the angle at which the fraction of electrons located within this region is maximized.

\begin{figure*}[t!]
	\centering
	\includegraphics[width=\textwidth]{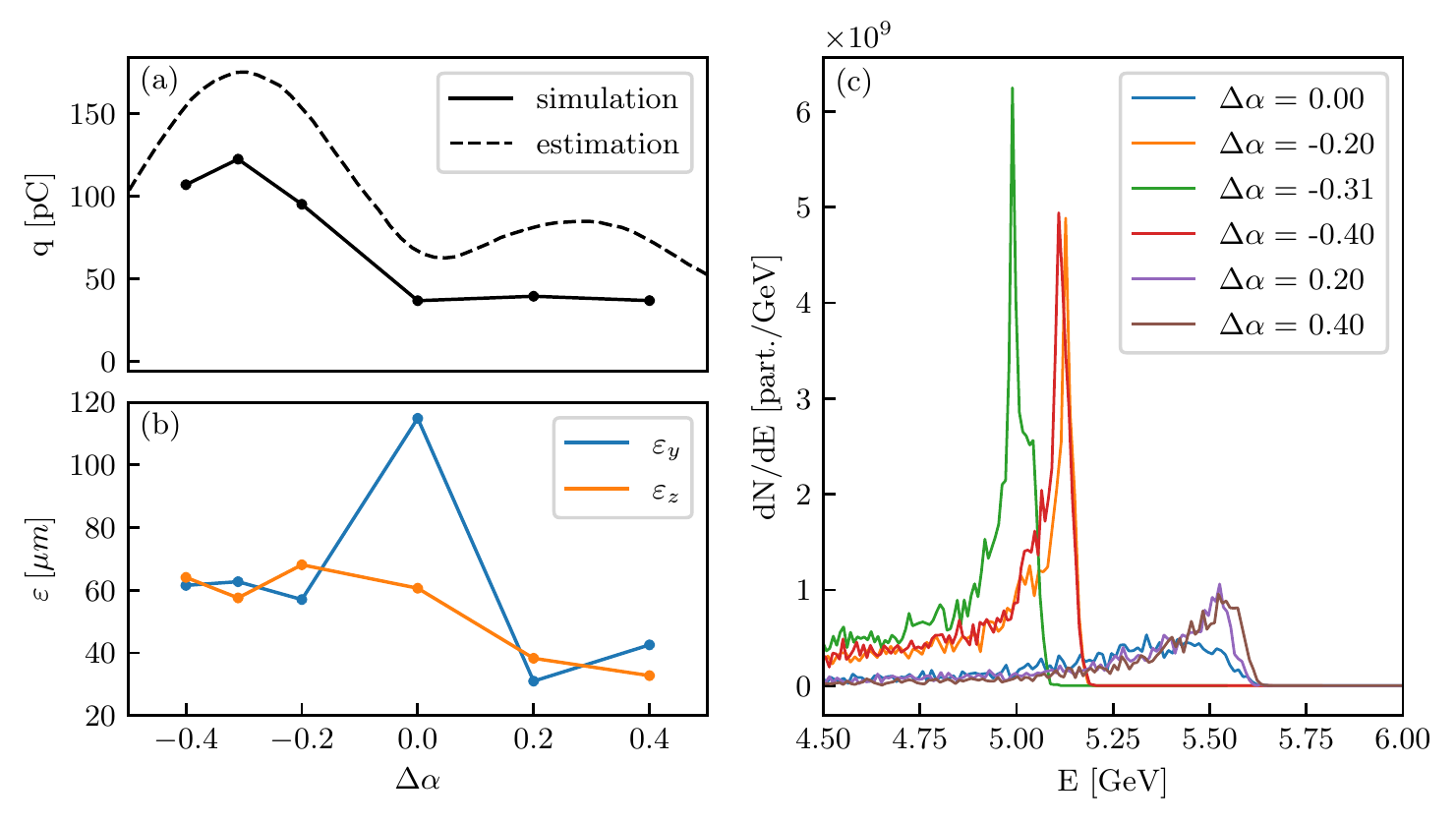}
	\caption{Simulation results for different rotation angles and laser pulse energy of 15 mJ. (a) Trapped charge depending on rotation angle. Dashed line gives the theoretical prediction, dots are the simulation results. (b) Emittance along $y$ and $z$ directions. (c) Energy spectra after 10 m of acceleration.}\label{fig4-rotate}
\end{figure*}

\begin{figure}[t!]
	\centering
	\includegraphics[width=\linewidth]{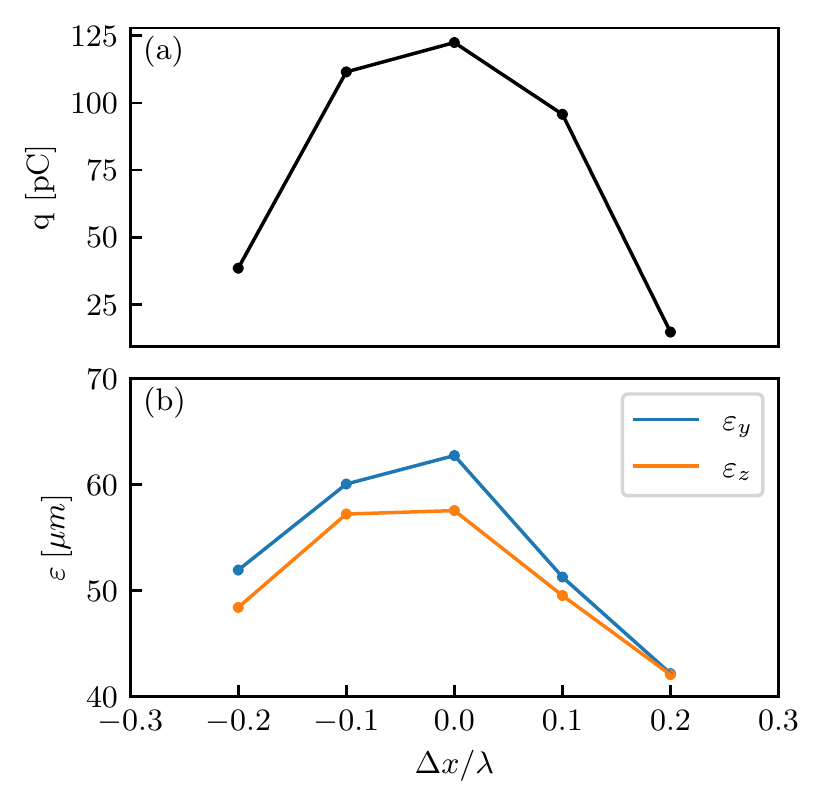}
	\caption{(a) Trapped charge depending on longitudinal shift. (b) Emittance along $y$ and $z$ directions.}\label{fig5-shift}
\end{figure}

In order to find the trapping condition we begin with the simple Hamiltonian for a test electron moving in 1d linear wakefield.
\begin{equation}
	H = \sqrt{1 + p_x^2} - \beta_bp_x  - \phi(\xi),
\end{equation}
where $\beta_b$ is the beta factor of the driver beam, $\xi = x - \beta_b t$, $\phi(\xi) = \phi_0\cos\xi$, $\phi_0$ is the amplitude of the wakefield. All the variables are measured in plasma units: the time is given in $\omega_p^{-1}$, where $\omega_p = \sqrt{4\pi n_0 e^2/m_e}$ is the plasma frequency, $n_0$ is the electron density, the coordinates are normalized to $c\omega_p^{-1}$, the particle energies are given in $m_ec^2$, the momenta are normalized to $m_ec$, the fields are measured in $m_ec\omega_p/e$. The point $\xi = 0$ corresponds to the zero of the longitudinal electric field after the acceleration phase $\xi\in [-\pi,\:0]$ (Fig.~\ref{fig3-phase}(a)).

An electrons with the initial conditions $\xi(0) = 0$, $p_x(0) = p_{0x}$ is trapped if its trajectory in the phase space lies within the outer separatrix shown in Fig.~\ref{fig3-phase}(b) as the dashed black line. This trajectory has the turning points at $\xi=\pm \pi$. The trapping condition can be written as follows:
\begin{equation}
	\gamma_0 - \beta_bp_{0x} < \gamma_b^{-1} + 2\phi_0.
\end{equation}
where $\gamma_0 = \sqrt{1+p_{0x}^2}$, $\gamma_b = (1-\beta_b^2)^{-1/2}$.

For the 2d case the trapping problem cannot be solved analytically. However, one can assume that the focusing region ($\xi\in[-\pi/2,\:\pi/2]$, Fig.~\ref{fig3-phase}.(a)) limits the trapping. Then, the new separatrix is a trajectory with turning points at the boundary of the focusing region ($\xi = \pm\pi/2$, Fig.~\ref{fig3-phase}(b), solid black line). Only particles moving inside the focusing domain are trapped. Thus, the 2d trapping condition can be written as follows: 
\begin{equation}\label{eq:2d_px}
	\gamma_0 - \beta_bp_{0x} < \gamma_b^{-1}+\phi_0.
\end{equation}
Assuming the transverse momentum $p_r$ to be much smaller than the longitudinal one, we can add $p_r$ into Eq. \eqref{eq:2d_px}: $\gamma_0 =  \sqrt{1+p_{0x}^2+p_{0r}^2}$, then we obtain the following trapping condition:
\begin{equation}\label{eq:trap}
	\frac{(p_{x 0} - p_c)^2}{a^2} + \frac{p_{r 0}^2}{b^2} < 1,
\end{equation}
where $p_c = \gamma_b\beta_b T$, $a^2 = \gamma_b^2(\gamma_b^2 T^2 - 1)$, $b^2 = \gamma_b^2 T^2 - 1$, $T = \gamma_b^{-1} + \phi_0$.
This region in shown in Fig.~\ref{fig2-en_ang}(b) with blue shading. Here, the electron distribution has been rotated at the optimal angle $\Delta\alpha = -0.31$ rad.

\section{Acceleration simulations}\label{sec:simulation}

We preformed a series of 3d PIC simulations with VLPL code to determine the trapped charge, emittance and energy spectrum for different injection parameters. 
The moving window with the sizes $L_x\times L_y\times L_z = 4\lambda_p\times 4\lambda_p\times4\lambda_p$ was sampled by a grid with the step sizes $h_x\times h_y\times h_z = 0.01\lambda_p\times0.02\lambda_p\times0.02\lambda_p$, where $\lambda_p = \sqrt{\pi m_e c^2/(n_pe^2)}=1.26~\rm{mm}$ is the plasma wavelength corresponding to the electron density $n_p = 7\cdot 10^{14}\:cm^{-3}$. These parameters are close to those of the AWAKE experiment. The simulated acceleration distance along the $x-$axis is 8000 $\lambda_p$ ($\sim$10 m).

We do not simulate here the self-modulation of the long proton driver used in the AWAKE experiments. Rather, we employ the so called "toy model" \cite{toy-model}, where the driver is presented as a short rigid proton bunch with $\gamma_b = 420$. It has the Gaussian shape with the longitudinal and transverse sizes $\sigma_z = \sigma_r = 0.17\,\lambda_p$. This driver excites a quasi-linear wakefield with the amplitude of $E_x \sim 0.24 E_0$ on the axis, where $E_0 = m_ec\omega_p/e$ is the wave breaking limit.

\begin{figure*}[t!]
	\centering
	\includegraphics[width=\textwidth]{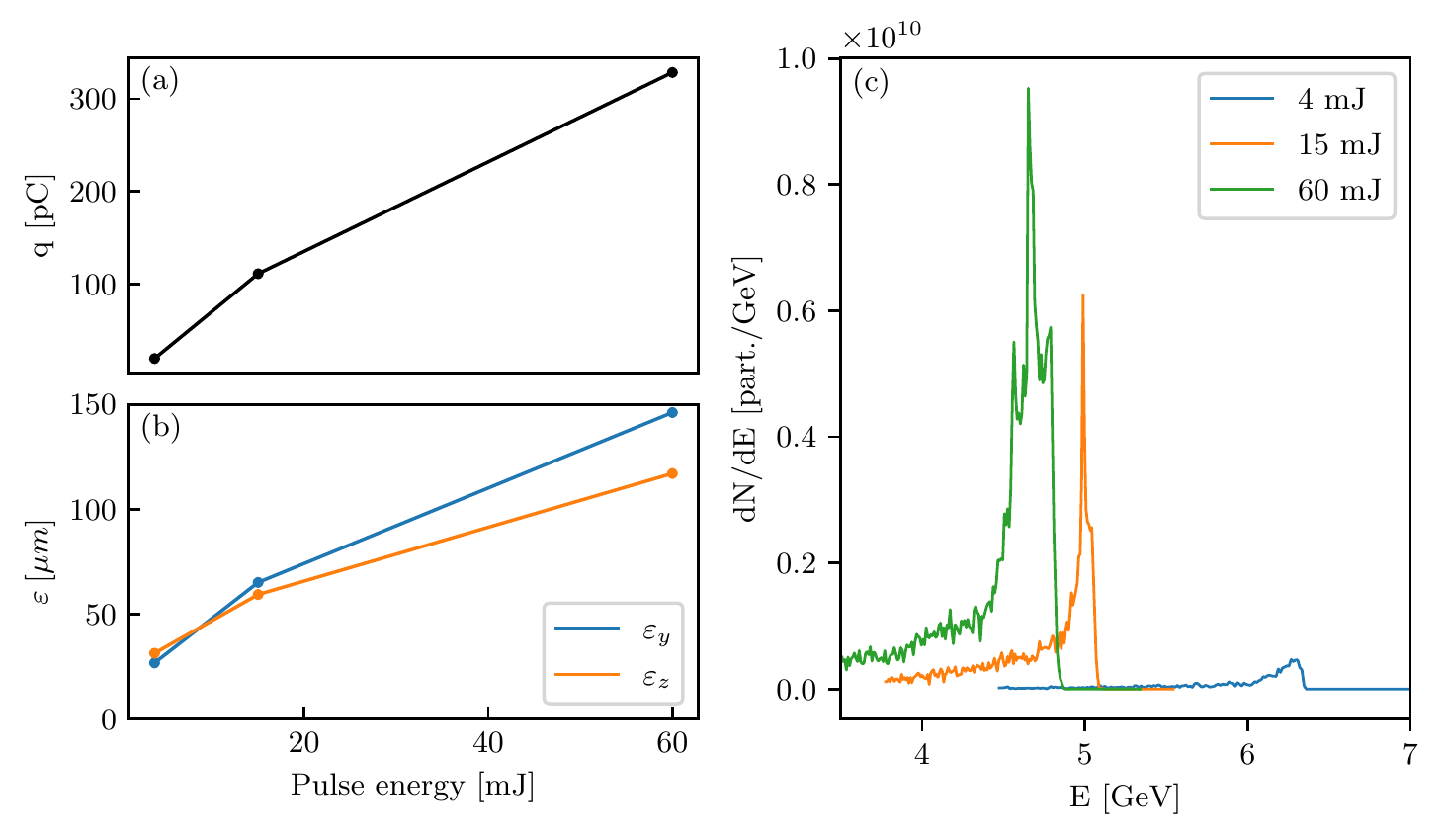}
	\caption{(a) Trapped charge depending on laser pulse energy. (b) Emittance along $y$ and $z$ directions. (c) Energy spectra for different laser pulse energies.}\label{fig6-a0}
\end{figure*}

In the first series of simulations, the electrons were injected at the axis, at the phase of zero $E_x$ field and we varied the rotation angle $\Delta \alpha$. The results are presented in Fig.~\ref{fig4-rotate}. The dashed line in Fig.~\ref{fig4-rotate}(a) corresponds to the estimated trapped charge  according to the condition \eqref{eq:trap}. The black dots show simulation results after $10~$m of acceleration. The simulation points do not coincide quantitatively with the prediction and we did not expect that, because the theory is oversimplified: the wake is assumed to be linear, the transverse field is not taken into account. Yet, the two-hill trend and the positions of maxima are well predicted. Also it turned out that the initial target orientation is not optimal with respect to the emittance (Fig.~\ref{fig4-rotate}(b)). The asymmetry is the largest here, $\varepsilon_y$ is twice as big as $\varepsilon_z$. The electron energy spectra demonstrate spiky profiles for charges $\gtrsim100~$pC with the energy $\sim5~$GeV and spread $\lesssim 1$\%.

In the second series of simulations, the electrons were injected into different phases of the wakefield. The phase of zero longitudinal field before the accelerating region optimizes the trapped charge.  Any deviation from it lowers the number of trapped electrons, see Fig.~\ref{fig5-shift}(a). The emittance also decreases and seems to be correlated with the number of particles (Fig.~\ref{fig5-shift}(b)).

In the third series of simulations, we varied the laser pulse energy while keeping the optimal injection phase and angle. For the laser pulse energy of 60 mJ, the trapped charge increases up to over 300 pC (Fig.~\ref{fig6-a0}(a)). However, simultaneously, the emittance experiences a significant growth, and the asymmetry between the $y$ and $z$ directions grows as well (Fig.~\ref{fig6-a0}(b)). It is caused by the trapping of large amount of low-energy uncollimated particles generated by the high energy laser pulse. These electrons still satisfy the trapping condition, so that the rotation fixes the asymmetry less effectively. The energy spectra (Fig.~\ref{fig6-a0}(c)) are mono-energetic with the relative spread $\sim 1$\% and the peak energy of 4-6 GeV for charges over $\sim$ 100 pC  after 10 m.

\section{Conclusion}\label{sec:conclusion}
Summarizing, in this paper we numerically studied a new laser-assisted injection scheme for PWFA. We presented a series of two-step simulations. In the fist simulation step, MeV electrons are generated from laser-solid interaction, in the second step, these electrons are injected into a axisymmetric quasilinear plasma wakefield with parameters close to those of the AWAKE experiment. The evolution of the trapped electron bunch is tracked for 10 m. It was established that the optimal injection point is the phase of zero longitudinal field on the axis before the accelerating phase. An important parameter is the angle at which particles are injected into a wake. It strongly influences the trapped charge and transverse symmetry of obtained electron bunch. An approximate trapping condition in a two-dimensional wakefield was derived in order to optimize the witness charge. Simulations showed that the positions of maxima and general trend for witness charge dependence on the rotation angle are well predicted by this 2d trapping condition. The optimization of injection angle with respect to the initial target position, when the direction of reflected laser pulse and driver velocity coincide, improves the witness bunch parameters significantly. The witness charge increases by a factor of three, while the transverse normalized emittance reduces by half, improving the transverse symmetry. The energy spectra demonstrate spiky profiles with the reached energies  $\sim$~5~GeV and relative spread $\lesssim$ 1\% for trapped charges over 100 pC. The increase in laser pulse energy is accompanied by slower than linear increase in the trapped charge. At this point, the upper limit of beam loading is reached and a further significant improvement of the witness parameters is unlikely. At the same time, the transverse emittance also experiences a significant growth due to the appearance of large amount of low energetic, undirected particles in the initial angular distribution, which however still fit the trapping condition. Thus, the optimal laser pulse energy is in a few 10s of mJ.

The laser-plasma interaction in the injection scheme discussed above could be optimized further following the recent experimental findings  \cite{mev-e-new}. This is the subject of on-going work.

\section*{Acknowledgments}
This work has been supported by DFG grant PU 213/6-2.


\begin{thebibliography}{99}
	\bibitem{RMP81-1229}
	E. Esarey, C. B. Schroeder, and W. P. Leemans,
	Rev. Mod. Phys. \textbf{81}, 1229 (2009).
	\bibitem{RAST9-63}
	M.J.Hogan,
	Reviews of Accelerator Science and Technology \textbf{9}, 63 (2016).
	\bibitem{RAST9-85}
	E. Adli and P. Muggli,
	Reviews of Accelerator Science and Technology \textbf{9}, 85 (2016).
	\bibitem{RMP90-035002}
	M.C. Downer, R. Zgadzaj, A. Debus, U. Schramm, and M.C. Kaluza,
	Rev. Mod. Phys. 90, 035002 (2018).
	\bibitem{inj-ext1} H. Suk, N. Barov, J. B. Rosenzweig, and E. Esarey, Phys. Rev. Lett. 86, 1011 (2001)
	\bibitem{inj-ext2} M. J. H. Luttikhof, A. G. Khachatryan, F. A. van Goor, and K.-J. Boller, Phys. Plasmas 14(8), 083101 (2007)
	\bibitem{inj-dens1} H. Suk, N. Barov, J. B. Rosenzweig, and E. Esarey, Phys. Rev. Lett. 86, 1011 (2001)
	\bibitem{inj-dens2} X. L. Xu, F. Li, W. An, T. N. Dalichaouch, P. Yu, W. Lu, C. Joshi and W. B. Mori, Phys. Rev. Accel. Beams 20, 111303 (2017)	 
	\bibitem{inj-dens3} A. Martinez de la Ossa, Z. Hu, M. J. V. Streeter, T. J. Mehrling, O. Kononenko, B. Sheeran and J. Osterhoff, Phys. Rev. Accel. Beams 20, 091301 (2017)
	
	\bibitem{inj-ion1} E. Oz, S. Deng, T. Katsouleas, P. Muggli, C. D. Barnes, I. Blumenfeld, F.
	J. Decker, P. Emma, M. J. Hogan, R. Ischebeck, R. H. Iverson, N. Kirby,
	P. Krejcik, C. O’Connell, R. H. Siemann, D. Walz, D. Auerbach, C. E.
	Clayton, C. Huang, D. K. Johnson, C. Joshi, W. Lu, K. A. Marsh, W. B.
	Mori, and M. Zhou, Phys. Rev. Lett. 98, 084801 (2007).
	\bibitem{inj-ion2} N. Kirby, I. Blumenfeld, C. E. Clayton, F. J. Decker, M. J. Hogan, C.
	Huang, R. Ischebeck, R. H. Iverson, C. Joshi, T. Katsouleas, W. Lu, K. A.
	Marsh, S. F. Martins, W. B. Mori, P. Muggli, E. Oz, R. H. Siemann, D. R.
	Walz, and M. Zhou, Phys. Rev. ST Accel. Beams 12, 051302 (2009).
	\bibitem{trojan} B. Hidding, G. Pretzler, J. B. Rosenzweig, T. Konigstein, D. Schiller, and D. L. Bruhwiler
	Phys. Rev. Lett. 108, 035001 (2012).
	
	\bibitem{mev-e} I. Tsymbalov, D. Gorlova, S. Shulyapov, V. Prokudin, A. Zavorotny, K. Ivanov, R. Volkov, V. Bychenkov, V. Nedorezov, A. Paskhalov, N. Eremin and A. Savel’ev, Plasma Phys. Control. Fusion 61 (2019) 075016.
	\bibitem{awake}
	E. Adli, A. Ahuja, O. Apsimon, R. Apsimon, A. M. Bachmann, D. Barrientos, F. Batsch, J. Bauche, V. K. Berglyd Olsen, M. Bernardini, T. Bohl, C. Bracco, F. Braunmuller, G. Burt, B. Buttenschon, A. Caldwell, M. Cascella, J. Chappell, E. Chevallay, M. Chung, D. Cooke, H. Damerau, L. Deacon, L. H. Deubner, A. Dexter, S. Doebert, J. Farmer, V. N. Fedosseev,R. Fiorito, R. A. Fonseca, F. Friebel, L. Garolfi, S. Gessner, I. Gorgisyan, A. A. Gorn, E. Granados, O. Grulke, E. Gschwendtner, J. Hansen, A. Helm, J. R. Henderson, M. Huther, M. Ibison, L. Jensen, S. Jolly, F. Keeble, S. Y. Kim, F. Kraus, Y. Li, S. Liu, N. Lopes, K. V. Lotov, L. Maricalva Brun, M. Martyanov, S. Mazzoni, D. Medina Godoy, V. A. Minakov, J. Mitchell, J. C. Molendijk, J. T. Moody, M. Moreira, P. Muggli, E. Oz, C. Pasquino, A. Pardons, F. Pena Asmus, K. Pepitone, A. Perera, A. Petrenko, S. Pitman, A. Pukhov, S. Rey, K. Rieger, H. Ruhl, J. S. Schmidt, I. A. Shalimova, P. Sherwood, L. O. Silva, L. Soby, A. P. Sosedkin, R. Speroni, R. I. Spitsyn, P. V. Tuev, M. Turner, F. Velotti, L. Verra, V. A. Verzilov, J. Vieira, C. P. Welsch, B. Williamson, M. Wing, B. Woolley, and G. Xia, Nature 561, 363 (2018).
	
	\bibitem{toy-model} V. K. Berglyd Olsen, E. Adli, and P. Muggli, Phys. Rev. Accel. Beams 21, 011301 (2018).
	
	\bibitem{vlpl} A. Pukhov, Journal of plasma physics 61, 425 (1999).
	\bibitem{dla} A. Pukhov, Z.-M. Sheng and J. Meyer-ter-vehn, Phys. Plasmas 6, 2847–54 (1999)	
	\bibitem{mev-e-new} I. Tsymbalov, D. Gorlova, K. Ivanov, S. Shulyapov, V. Prokudin,	A. Zavorotny, R. Volkov, V. Bychenkov, V. Nedorezov and A Savel’ev, Plasma Phys. Control. Fusion 63 (2021) 022001
\end{thebibliography}
\end{document}